\documentclass[a4paper,12pt]{statsoc}
\usepackage{natbib}
\usepackage{geometry}
\usepackage{graphicx}
\usepackage{array}
\usepackage{amssymb,amsfonts,amsmath}
\usepackage{Sweave}
\usepackage{float}
\title[McGLM]{Multivariate Covariance Generalized Linear Models}

\author[Bonat, W.H and J\o rgensen, B.]{Wagner Hugo Bonat}
\address{Department of Statistics, Paran\'a Federal University, Curitiba, Brazil. \\
Department of Mathematics and Computer Science, University of Southern Denmark, Odense, Denmark.}
\email{wbonat@ufpr.br}
\author[Bonat, W.H and J\o rgensen, B.]{Bent J\o rgensen}
\address{Department of Mathematics and Computer Science, University of Southern Denmark, Odense, Denmark.}

\begin{document}

\begin{abstract}
We propose a general framework for non-normal multivariate data analysis called multivariate covariance generalized linear models (McGLMs), designed to handle multivariate response variables, along with a wide range of temporal and spatial correlation structures defined in terms of a covariance link function combined with a matrix linear predictor involving known matrices. The method is motivated by three data examples that are not easily handled by existing methods. The first example concerns multivariate count data, the second involves response variables of mixed types, combined with repeated measures and longitudinal structures, and the third involves a spatio-temporal analysis of rainfall data. The models take non-normality into account in the conventional way by means of a variance function, and the mean structure is modelled by means of a link function and a linear predictor. The models are fitted using an efficient Newton scoring algorithm based on quasi-likelihood and Pearson estimating functions, using only second-moment assumptions. This provides a unified approach to a wide variety of different types of response variables and covariance structures, including multivariate extensions of repeated measures, time series, longitudinal, spatial and spatio-temporal structures.
\end{abstract}

\section{Introduction}
	The analysis of non-normal multivariate data currently involves a choice between a considerable array of different modelling frameworks, ranging from, say, generalized estimating equations (GEE) and time-series models to generalized linear mixed models and model-based geostatistics. Each framework allows the modelling of a specific type of dependence or correlation structure, without fitting into any clear overall pattern. Current software implementations have, as we shall see below, limited capacity in terms of the complexity and size of data that can be handled.

This situation stands in sharp contrast to the univariate case, where Nelder and Wedderburn's (1972) \nocite{Nelder:1972} generalized linear models (GLMs) provide a unified and versatile approach to regression modelling of normal and non-normal data, implemented in an efficient fitting algorithm. A further advantage of the GLM approach is that estimation and inference for GLMs require only second-moment assumptions.

In order to obtain a multivariate modelling framework of comparable range and versatility, we shall propose the class of multivariate covariance generalized linear models (McGLMs), which, following \citet{Pourahmadi:1999}, are specified via separate link functions and linear predictors for the mean vector and covariance matrix, respectively. This allows a unified approach to analysis of multivariate correlated data, taking into account response variable of mixed types, and allowing a wide range of covariance structures for repeated measures, longitudinal, spatial and spatio-temporal data. The models are fitted by means of quasi-likelihood and Pearson estimating functions, based on second-moment assumptions, and implemented in an efficient Newton scoring algorithm.

The idea of modelling a function of the covariance matrix by a linear structure goes back at least as far as \citet{Anderson:1973}, followed later by \citet{Chiu:1996}, who used the matrix logarithm as covariance link function. More recently, the idea was extended in several different ways by \citet{Pourahmadi:1999, Pourahmadi:2011}, \citet{Pan:2003} and \citet{Zhang:2015}, among others. These authors consider mainly the multivariate normal distribution, whereas we shall use a variance function to take non-normality into account in the style of \citet{Liang:1986}. Contrary to the latter authors we shall, however, emphasize the need to model the covariance structure explicitly, rather than treating it as a nuisance parameter.

The availability of standard software is an indicator for which kind of statistical methods are in currently use by the statistical and scientific communities. It is hence interesting to note that well-established \texttt{R} packages such as \texttt{lme4} \citep{lme4:2014} and \texttt{nlme} \citep{nlme:2013} do not deal with multivariate response variables. In the Bayesian context the flexible packages \texttt{INLA} \citep{INLA:2014} and \texttt{MCMCpack} \citep{MCMCpack:2011} do not deal with multivariate response variables, judging from the package documentation. In \texttt{R}, there are at least two generalized linear mixed models packages that can deal with multivariate response variables, namely \texttt{MCMCglmm} \citep{MCMCglmm:2010}, which uses Monte Carlo Markov Chain (MCMC) methods in the Bayesian framework, and the package \texttt{SabreR} \citep{sabreR:2010}, which uses marginal likelihood, but is limited to dealing with at most three response variables. The modelling of the covariance structure is currently restricted to making a selecting from a short list of pre-specified covariance structures, such as autoregression or compound symmetry. We were not able to find any \texttt{R} packages for fitting joint mean-covariance models, not even in the multivariate normal case. In \texttt{SAS} the \texttt{GLIMMIX} procedure for generalized linear mixed models (GLMMs) deals with multivariate response variables, but is limited to the exponential family of distributions and a few pre-determined covariance structures \citep{SAS}. Other software platforms for fitting generic random effects models via MCMC, such as \texttt{JAGS} \citep{Plummer03jags:a} or \texttt{WinBUGS} \citep{Lunn2000}, can deal with multivariate response variables, but carry substantial overheads in terms of computational times and convergence checks, while being restricted to a small set of pre-specified covariance structures and probability distributions. These limitations on current software availability for joint mean-covariance modelling of multivariate response variables may reflect either a lack of interest on the part of software users, or a lack of sufficiently flexible modelling frameworks. In any case, we will use the latter as motivation for developing the new class of McGLMs. 

We now present three correlated data examples along with a short review of currently available methods for each type of data. The examples were selected in order to highlight some of the limitations of current methodology, while illustrating the range of different problems that may be handled by the McGLM method. 
    


		\subsection{Data set 1: Australian health survey}\label{data1}
		The first data set is from the Australian Health Survey for 1987--1988 \citep{Partha:1997,Cameron:1998}. We selected the following five count response variables for our analysis: number of consultations with a doctor or specialist ($\texttt{Ndoc}$) or with health professionals ($\texttt{Nndoc}$); total number of prescribed and non prescribed medications used in the past two days ($\texttt{Nmed}$); number of nights in a hospital during the most recent admission ($\texttt{Nhosp}$) and number of admissions to a hospital, psychiatric hospital, nursing or convalescence home in the past $12$ months ($\texttt{Nadm}$). The data set had nine covariates concerning social conditions (see Appendix for details). There were $5190$ respondents and no missing data.

\setkeys{Gin}{width=0.99\textwidth}
\begin{figure}[htbp]
\centering
\includegraphics{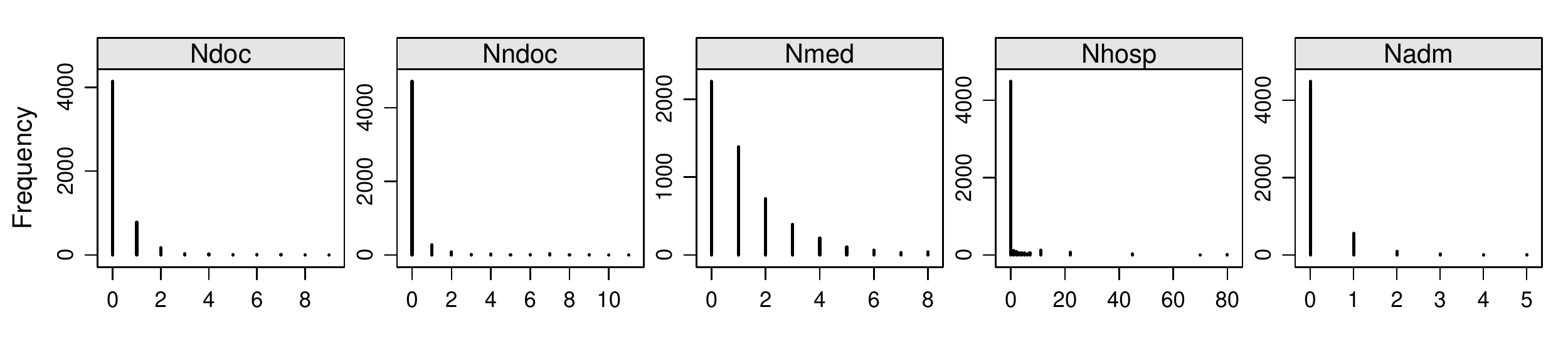}
\caption{Histograms for each outcome variable of the Australian health survey data.}
\label{fig:exploratory1}
\end{figure}

This example illustrates the fairly common situation of a multivariate regression problem with non-normal (discrete) response variables. The histograms in Figure \ref{fig:exploratory1} suggest that the five error distributions may not be identical, and hint at potential problems with excess of zeroes and under/overdispersion. These problems may, in turn, reflect on the solution to the main questions of the analysis, namely assessing the effects of the covariates on each outcome, and determining the residual correlation structure. 

Given currently available software, it is a daunting task to select a suitable marginal error distribution for each of the five response variables. Besides the classical Poisson and negative binomial distributions, other distributions such as the Neyman Type A \citep{Dobbie:2001} or the Poisson-inverse Gaussian (PIG) \citep{Holla:1967} may be relevant. Different distributions may have to be fitted by separate software packages, each of which comes with its own set of problems due to badly behaved likelihood function etc. 

If we decide to use formal methods of model selection, we are faced with the choice of selection criterion, such as the Akaike or Bayesian information criterion in the likelihood framework, or the deviance information criterion in the Bayesian framework. The Bayesian case involves additional work due to the need for choosing suitable prior distributions. These problems persist in the special case where all error distributions belong to the same family. One option is the multivariate Poisson regression \citep{Tsionas:2007}, which is suitable for multivariate count data, but is restricted to positive correlations and equidispersed data. A second option is the multivariate negative binomial distribution proposed by \citet{Shi:2014}. Such models are not easy to fit, and require careful attention to the implementation of algorithms and starting values. The assumption of a common error distribution required for these models may, however, not be satisfied in practice, and methods for handling the case of unequal marginal distributions do not seem to be easily available.

A different approach for correlated data is the family of generalized linear mixed models (GLMM) \citep{Breslow:1993,Fong:2010}, which is based on specifying a GLM conditionally on a multivariate latent distribution, often the multivariate normal. A specific example of a GLMM for multivariate count data was presented by \citet{Motta:2013}. GLMMs are computationally demanding, and many different algorithms have been proposed in the past three decades, see \citet{McCullogh:1997} and \citet{Fong:2010} for reviews and further references. 

A further aspect of GLMMs that gives rise to concern is the general lack of a closed-form expression for the likelihood and the marginal distribution of the data vector. This makes model selection even more complicated than for the marginal models discussed above. A related question is the special interpretation of parameters inherent from the construction of GLMMs. Thus, the covariate effects are conditional on the latent variables, whereas the correlation structure is marginal for the latent variables rather than for the response variables. An interesting discussion of random-effects and marginal models may be found in \citet{Lee:2004}.

Additional methods for specifying models for multivariate response variables include the copula models \citep{Krupskii:2013} and the class of hierarchical generalized linear models \citep{Lee:1996}. The fact that several different approaches are available for multivariate regression modelling, none of which is particularly easy to use, amplifies our call for a universal multivariate modelling framework, preferably one that facilitates model selection and allows marginal interpretation of parameters.

		\subsection{Data set 2: Respiratory physiotherapy on premature newborns}\label{data2}
		We consider some aspects of a prospective study to assess the effect of respiratory physiotherapy on the cardiopulmonary function of ventilated preterm newborn infants with birth weight lower than 1500 g. The study had three response variables: respiratory rate (\texttt{RR}), heart rate (\texttt{HR}) and oxygen saturation (\texttt{O}$_2$\texttt{Sat}). The \texttt{HR} and \texttt{O}$_2$\texttt{Sat} data were collected by electronic monitoring and \texttt{RR} by means of a stopwatch.  Response variables were taken three times: before starting the physiotherapy (\texttt{Evaluation 1}), immediately after finishing (\texttt{Evaluation 2}), and five minutes after finishing the physiotherapy (\texttt{Evaluation 3}). Sixteen newborns were evaluated in consecutive sessions by two therapists at the neonatal unit. The number of evaluation days varied between $2$ and $7$ days. The data set has $16$ covariates concerning health conditions and there are $270$ cases (see the Appendix). Figure \ref{fig:exploratory2} shows the individual and average trajectories by outcome and evaluation.

\setkeys{Gin}{width=0.99\textwidth}
\begin{figure}[t]
\centering
\includegraphics{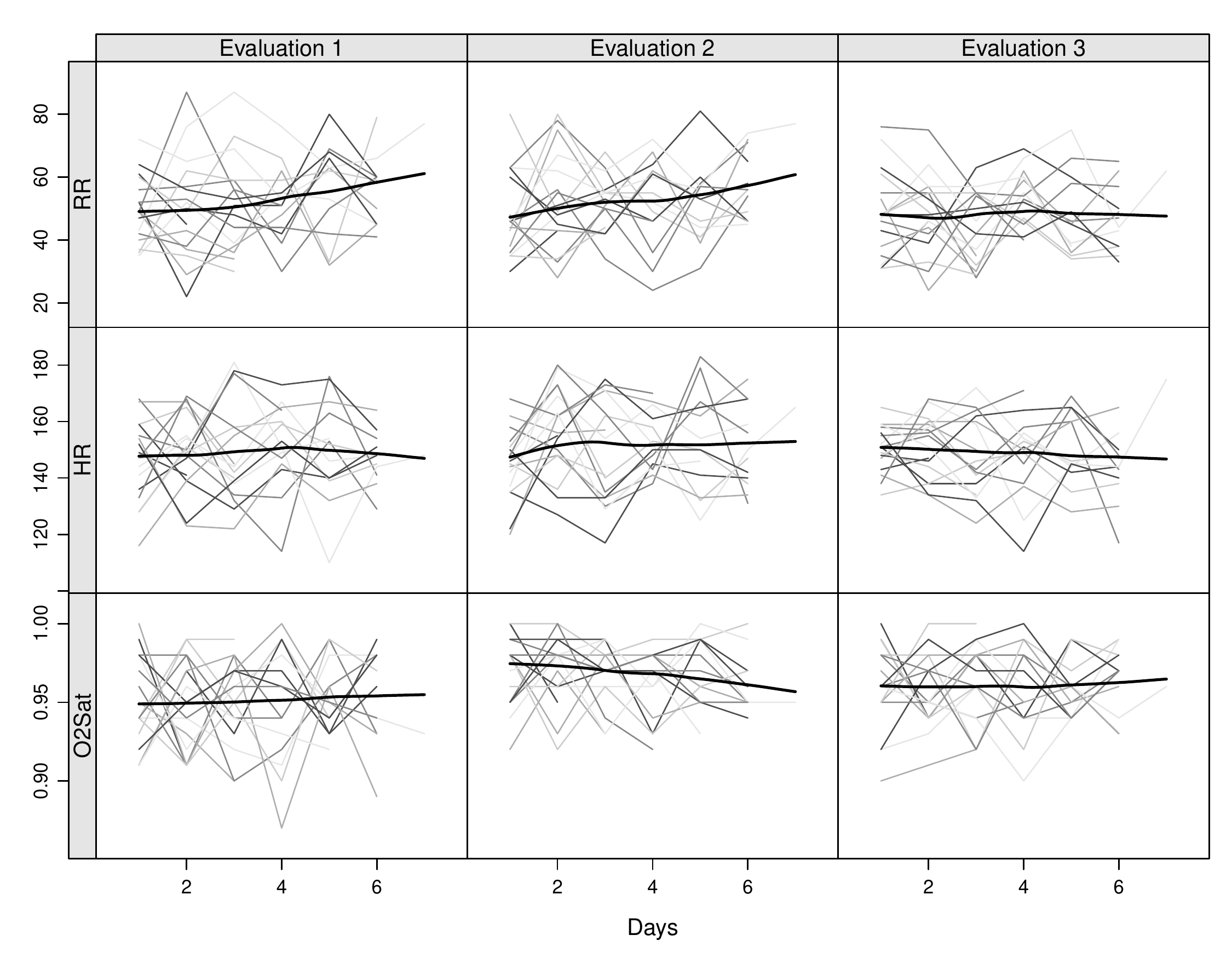}
\caption{Individual and average (solid line) trajectories by outcome and evaluation for the Respiratory physiotherapy data.}
\label{fig:exploratory2}
\end{figure}

The main goal of the investigation was to assess the effect of respiratory physiotherapy on the outcome variables, while taking into account the effects of covariates and the correlation induced by the repeated measures and the longitudinal structures. A special feature of these data is that the outcome variables are of mixed types. Thus, the variables \texttt{HR} and \texttt{RR} are continuous, whereas the oxygen saturation variable \texttt{O}$_2$\texttt{Sat} takes values in the unit interval, including about $5\%$ exact ones, making it hard to propose a suitable probability distribution for this variable. We may, of course, use for example the beta \citep{Bonat:2015} or the simplex distribution \citep{Zhang:2014} with some ad hoc method for dealing with the exact ones. A better option may be to use the beta distribution inflated with ones \citep{Raydonal:2010}, but this model is complicated to fit and interpret. It may hence be preferable in this situation to use a quasi-likelihood method based on second-moment assumptions, which is easier to fit and interpret.

Similar to what we saw in Example $1$, the literature may be divided into two main approaches: marginal models, mostly based on the GEE approach \citep{Brien:2004,Rochon:1996,Gray:2000}, and random-effects models based on GLMMs, see \citet{Verbeke:2014}. These authors also provide an extensive review of models for response variables of mixed type, whereas \citet{Fieuws:2007} reviewed random-effects models for multivariate repeated measures. The question of how to model the covariance structure for repeated measures and longitudinal data is often solved by choosing from a short list of options, such as compound symmetry, autoregressive, banded and unstructured \citep{Diggle:2002}. Such choices are, however, not suitable for the combination of repeated measures and longitudinal data found in the present data, thereby motivating the development of a more general and flexible approach for covariance modelling in multivariate data analysis. 

		\subsection{Data set 3: Venezuelan rainfall data}\label{data3}
		This example concerns monthly rainfall data from $80$ stations in the Venezuelan state of Gua\'arico for a period of $16$ years ($192$ months). The data set has $15360$ cases with $1092$ missing data. We also have the spatial coordinates (latitude and longitude) of the $80$ stations available, along with the covariate \texttt{height} (height above sea level). The data were previously analyzed by \cite{Sanso:1999} using Bayesian MCMC methods, based on a censored and transformed multivariate normal distribution. 

\begin{figure}[htbp]
\centering
\includegraphics{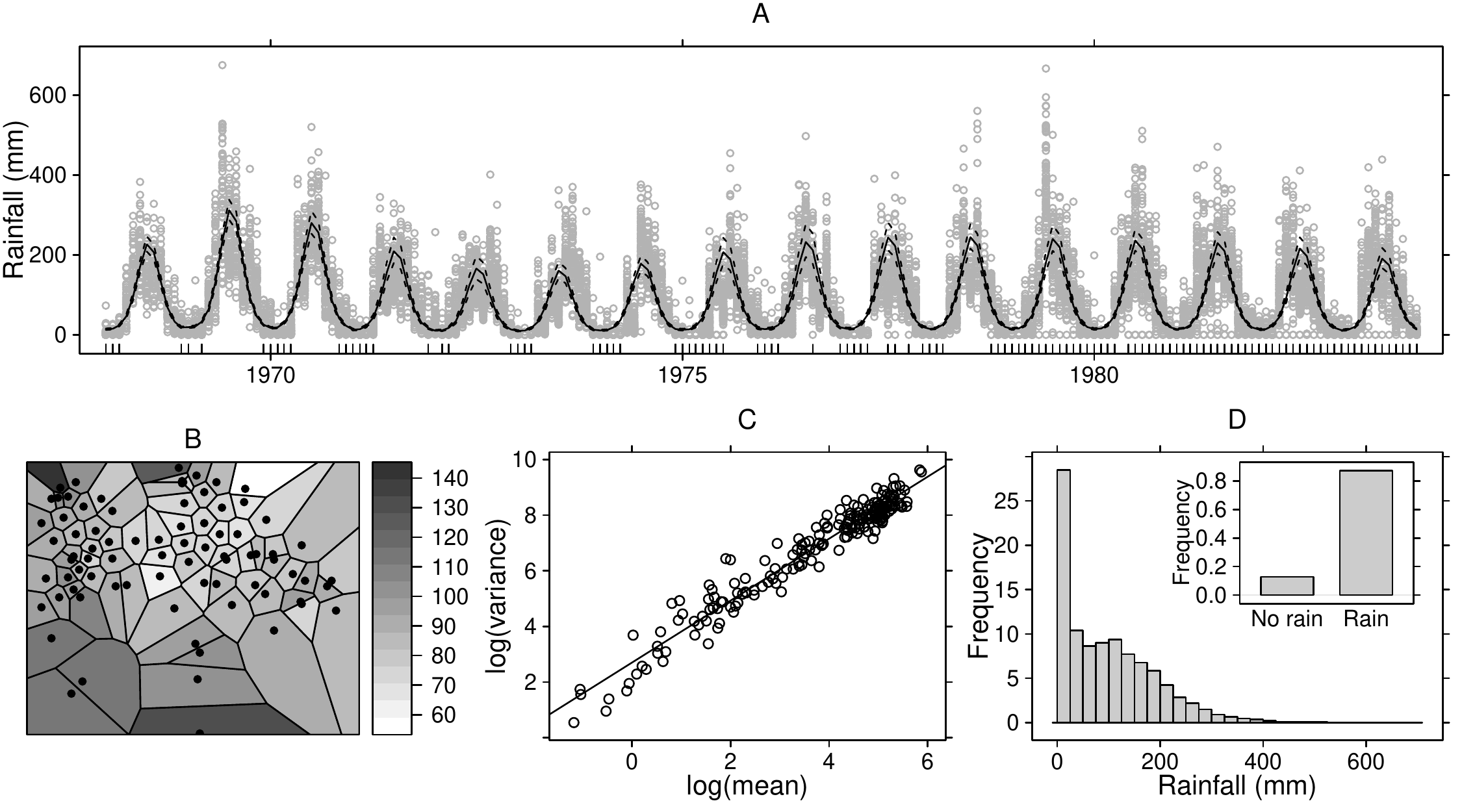}
\caption{Time series plot for Venezuelan rainfall data with fitted values (A). Voronoi tessellation map based on spatial coordinates, colored by the average monthly rainfall for the whole period (B). Taylor plot (spatial mean and variance in double logarithmic scale) (C). Histogram of monthly rainfall for the whole period (D).}
\label{fig:exploratory3}
\end{figure}

The statistical modelling of rainfall data involves a number of challenges, such as the need for simultaneous modelling of seasonal and geographical variation, the complicated nature of the spatio-temporal correlation structure, the special form of the marginal distribution (having a discrete component at zero), and the possible influence of the sampling scale on the form of the analysis \citep{Dunn:2004}. The plots shown in Figure \ref{fig:exploratory3} illustrate some of these features for the Venezuelan rainfall data. In particular, the histogram in panel D highlights the right-skewed distribution and the considerable proportion of exact zeroes (around $13\%$), whereas the approximate linearity of the Taylor plot in Panel C suggests a variance function of power form.

\setkeys{Gin}{width=0.99\textwidth}

A simple model for the marginal distribution of total rainfall $Y$ over a certain time period is to write $Y = R_1 + \cdots + R_N$, where $N$ is the number of rainfall episodes, assumed to be Poisson distributed, and the i.i.d. variables $R_i$ are the amounts of rain for each episode, with the convention $Y=0$ for $N=0$, corresponding to a discrete component at zero. A special case of this compound Poisson model is the Tweedie family \citep{Jorgensen1997b} (where the $R_i$ are gamma distributed), with power variance functions, in agreement with the Taylor plot of Figure \ref{fig:exploratory3}. The Tweedie model has been successfully applied to rainfall data by \citet{Dunn:2004} and \citet{Hasan2010,Hasan:2012}. These authors, however, assume independent data, which is not realistic for the present data set.

A popular approach for analyzing rainfall data \citep{Chandler:2002,Sigrist:2012} is to use separate models for the discrete component, indicating the number of wet periods, and the continuous component, indicating the amount of rain for wet periods \citep{Stern:1984,Wilks:1990}. A variety of distributions have been proposed for modelling the continuous component of rainfall under the independence assumption, including the log-normal, Weibull, generalized log-normal, gamma and mixed gamma distributions \citep{Hasan2010,Hasan:2012}. While these distributions may have their merits for analyzing rainfall data, the above compound Poisson model seems more natural, and the Tweedie family is flexible enough to mimic many of the shapes of other distributions.  

Turning now to the question of spatio-temporal modelling of rainfall data, one possibility is to use models based on marked point processes \citep{Wheater:2000,Cowpertwait:2006}, which may be useful for detailed simulation studies. Another approach is to follow the conventional geostatistical paradigm, assuming a parametric covariance function \citep{Diggle:2007}. There are several parametric families available for modelling the joint space-time covariance structure \citep{Cressie:1999,Gneiting:2002}, although there are issues with their interpretability and computational complexity, making it difficult to handle large data sets with this approach.

A different approach to spatio-temporal modelling is to take into account the fundamental difference between the spatial and temporal dimensions, the latter obeying a natural ordering which is not present in the spatial dimension. It may hence be natural to assume a dynamic temporal evolution model in combination with spatially correlated errors, see \citet{Sanso:1999,Sanso:2004,Sigrist:2012} and the monograph by \citet{Cressie:2011}. While providing a flexible form of spatio-temporal modelling, this method is also computationally demanding, and handles response variables with a discrete component at zero by means of a censored multivariate normal distribution, which does not provide as reasonable an interpretation as the Tweedie model.

A significant simplification may be obtained by assuming that the spatial domain is discrete, rather than being continuous as in the last two methodologies discussed above. This approach is used for example in disease mapping \citep{BESAG:1991}, where the covariance structure is determined by a neighborhood matrix. This is computationally less demanding, because for a given neighborhood structure we may specify the inverse covariance (or precision) matrix. The precision matrix, in turn, contains information about the structure of conditional independence of the data \citep{RUE+HELD:2005}. The proposed simplification may hence be seen as a reasonable compromise between model complexity and the capacity to model real data sets, achieveable by modelling the covariance structure using a linear combination of neighborhood matrices. To accommodate rainfall data, such a modelling strategy should allow for Tweedie distributed response variables with power variance functions. 

Section 2 presents the class of McGLMs, and Section 3 considers the Newton scoring algorithm. The three data examples presented here are analyzed in Section 4 using McGLMs. The results are discussed in Section 5, including some directions for future investigations.

\section{Multivariate covariance generalized linear models}\label{model}
	In this Section we will present the McGLM approach as an extension of GLMs. Let $\boldsymbol{Y}$ be an $N \times 1$ response vector, $\boldsymbol{X}$ an $N \times k$ design matrix and $\boldsymbol{\beta}$ a $k \times 1$ regression parameter vector. A GLM can be written in the following form:
\begin{eqnarray}
\label{modelGLM}
\mathrm{E}(\boldsymbol{Y}) &=& \boldsymbol{\mu} = g^{-1}(\boldsymbol{X} \boldsymbol{\beta})  \nonumber \\
\mathrm{Var}(\boldsymbol{Y}) &=& \boldsymbol{\Sigma} = \mathrm{V}(\boldsymbol{\mu};p)^{\frac{1}{2}}(\tau_0 \boldsymbol{I})\mathrm{V}(\boldsymbol{\mu};p)^{\frac{1}{2}}
\end{eqnarray}
where $g$ is the link function, $\mathrm{V}(\boldsymbol{\mu};p) = \mathrm{diag}(\vartheta(\boldsymbol{\mu};p))$, is a diagonal matrix whose main entries are given by the variance function $\vartheta(\cdot;p)$ applied elementwise to the vector $\boldsymbol{\mu}$. Finally $p$ and $\tau_0$ are the power and dispersion parameters, respectively, and $\boldsymbol{I}$ denotes the $N \times N$ identity matrix. 

The success enjoyed by the GLM framework comes from its ability to deal with a wide range of non-normal data using just two separate functions, namely the link and variance functions. The variance function plays an important role for GLMs, since different choices imply different assumptions about the response variable distribution. The power variance functions $\vartheta(\boldsymbol{\mu};p) = \mu^p$ are a frequent choice in the GLM framework. It characterizes the Tweedie family of distributions, whose most important special cases are the normal $(p=0)$, Poisson $(p=1)$, Gamma $(p=2)$ and inverse Gaussian $(p=3)$ distributions \citep{Jorgensen:1987, Jorgensen1997b}. But in spite of its flexibility the GLM approach has some limitations: it deals only with independent and univariate response variables, and the variance function is assumed to be known. 

Our main objectives are to extend the GLM approach to deal  with first non-independent data and second multivariate response variables. 
A third objective is to estimate the power parameter, which works as automatic model selection.

The Tweedie family is quite flexible for handling continuous response variables, but it is less flexible for discrete response variables. Therefore, we propose to use the Poisson-Tweedie family to deal with discrete data \citep{Shaarawi:2011}. The Poisson-Tweedie family has variance function $\vartheta(\boldsymbol{\mu};p) = \mu + \mu^p$, and many important models for count data are special cases, for example the Hermite $(p=0)$, Neyman Type A $(p=1)$, negative binomial $(p=2)$ and Poisson-inverse Gaussian $(p=3)$, see \citet{Jorgensen:2014}. When using the Poisson-Tweedie family, the matrix in (\ref{modelGLM}) takes the special form $\boldsymbol{\Sigma} = \mathrm{diag}(\boldsymbol{\mu}) + \mathrm{V}(\boldsymbol{\mu};p)^{\frac{1}{2}} ( \tau_0 \boldsymbol{I}) \mathrm{V}(\boldsymbol{\mu};p)^{\frac{1}{2}}$ because the dispersion parameter appears only in the second term. Another important case is when the response variable is binary, bounded, or the number of successes within a given number of trials. In that case the binomial variance function $\vartheta(\boldsymbol{\mu}) = \mu(1 - \mu)$ may be useful.

It is important to emphasize that by using just these three sets of variance functions we can deal with most frequently occurring types of response variables. Such flexibility is very useful, for example when analysing data set $1$, where the choice of count distribution for each response variable is not obvious. Using the Poisson-Tweedie variance function we can deal with zero-inflation and overdispersion, such as that observed in data set $1$. A similar situation appears for data set $2$, where we have a bounded response variable with exact ones, which can be well modelled using the binomial variance function. The Tweedie family, through its power variance function, can model zero-inflated and right skewed response variables, such as the monthly rainfall data.

In Eq. (\ref{modelGLM}) it is easy to see where the assumption of independent observations appears in the covariance matrix, which in turn suggests how to introduce dependence between observations. It is enough to change the identity matrix $\boldsymbol{I}$ to a non-diagonal matrix $\boldsymbol{\Omega}(\boldsymbol{\tau})$. This approach is similar to the idea of a \textit{working correlation} matrix in the Generalized Estimation Equation (GEE) framework \citep{Liang:1986, Zeger:1988}. Our approach differs from GEE in that we propose to model $\boldsymbol{\Omega}(\boldsymbol{\tau})$ in terms of a linear combination of known matrices, following the ideas of \citet{Anderson:1973} and \citet{Pourahmadi:2000}, i.e.
\begin{equation}
\label{linearcovariance}
h(\boldsymbol{\Omega}(\boldsymbol{\tau})) = \tau_0 Z_0 + \cdots + \tau_D Z_D.
\end{equation}
Here $h$ is the {\em covariance link function}, $Z_d$ with $d = 0, \ldots, D$ are known matrices reflecting the covariance structure, and $\boldsymbol{\tau} = (\tau_0, \ldots, \tau_D)$ is a $(D+1) \times 1$ parameter vector. This structure is a natural analogue of the linear predictor of the mean structure, and we call it a {\em matrix linear predictor}. Plugging the matrix linear predictor (\ref{linearcovariance}) into Eq. (\ref{modelGLM}), we obtain a so-called {\em covariance generalized linear model}.

Two new issues appear here, concerning how to specify the covariance link function $h$ and how to define the matrices $Z_d$. The first issue was discussed by \citet{Pinheiro:1996} and \citet{Pourahmadi:2011}. In this paper we will focus on well-known covariance link functions, such as the identity and the inverse functions. In Section \ref{results} we show how to specify the matrices $Z_d$ in order to obtain some well-known models for time series, spatial and space-time data. 

Many authors claim that a suitable covariance link function must provide an unrestricted and interpretable parametrization. While laudable, such a goal is probably over-optimistic, and does not seem to have been achieved yet, at least not for the general case \citep{Pourahmadi:2000,Pinheiro:1996}. The modified Cholesky decomposition proposed by \citet{Pourahmadi:2007} presents both features, but is restricted to the case where there is a natural ordering of the observations. In general, the identity and inverse covariance link functions allow for simple interpretations, 
but these covariance link functions do not provide unrestricted parametrizations. In fact it is quite hard to define the parameter space for $\boldsymbol{\tau}$. In Section \ref{estimation} we propose the so-called {\em reciprocal likelihood} algorithm where we use a tuning constant in order to control the step length of the algorithm and avoid unrealistic values for the parameter vector $\boldsymbol{\tau}$. From an algorithmic point of view, there is hence no need to require an unrestricted parametrization. 

The second main contribution of this paper is to extend the covariance generalized linear model to deal with multivariate response variables. Let $\mathbf{Y}_{N \times R} = \{\boldsymbol{Y}_1, \ldots, \boldsymbol{Y}_R\}$ be a response variable matrix and let $\mathbf{M}_{N \times R} = \{\boldsymbol{\mu}_1, \ldots, \boldsymbol{\mu}_R\}$ denote the corresponding matrix of expected values. To indicate that each response variable $\boldsymbol{Y}_r$ has its own covariance matrix we use the notation $\boldsymbol{\Sigma}_r = \mathrm{V}_r(\boldsymbol{\mu}_r;p)^{\frac{1}{2}} \boldsymbol{\Omega}_r(\boldsymbol{\tau}) \mathrm{V}_r(\boldsymbol{\mu}_r;p)^{\frac{1}{2}}$. It is important to emphasize that this matrix models the covariances within each response variable.
We introduce the $R \times R$ correlation matrix $\boldsymbol{\Sigma}_b$ to model the correlation between response variables. To specify the joint covariance matrix for all response variables, we adopt the generalized Kronecker product proposed by \citet{Martinez:2013} in the context of multivariate disease mapping. We hence define the McGLM by
\begin{eqnarray}
\label{McGLM}
\mathrm{E}(\mathbf{Y}) &=& \mathbf{M} = \{g_1^{-1}(\boldsymbol{X}_1 \boldsymbol{\beta}_1), \ldots, g_R^{-1}(\boldsymbol{X}_R \boldsymbol{\beta}_R)\} \nonumber    \\
\mathrm{Var}(\mathbf{Y}) &=& \boldsymbol{C} = \boldsymbol{\Sigma}_R \overset{G} \otimes \boldsymbol{\Sigma}_b 
\end{eqnarray}
where $\boldsymbol{\Sigma}_R \overset{G} \otimes \boldsymbol{\Sigma}_b = \mathrm{Bdiag}(\tilde{\boldsymbol{\Sigma}}_1, \ldots, \tilde{\boldsymbol{\Sigma}}_R)(\boldsymbol{\Sigma}_b \otimes \boldsymbol{I})\mathrm{Bdiag}(\tilde{\boldsymbol{\Sigma}}_1^T, \ldots, \tilde{\boldsymbol{\Sigma}}_R^T)$ is the generalized Kronecker product. The matrix $\tilde{\boldsymbol{\Sigma}}_r$ denotes the lower triangular matrix of the Cholesky decomposition of $\boldsymbol{\Sigma}_r$. The operator $\mathrm{Bdiag}$ denotes a block diagonal matrix and $\boldsymbol{I}$ denotes an $R \times R$ identity matrix.

\section{Estimation and inference}\label{estimation}
	In this Section we describe the estimating function approach used to estimate the model parameters \citep{Jorgensen:2004}. We divide the set of parameters into two subsets, $\boldsymbol{\theta} = (\boldsymbol{\beta}^\top, \boldsymbol{\lambda}^\top)^\top$. In this notation $\boldsymbol{\beta} = (\boldsymbol{\beta}_1^\top, \ldots, \boldsymbol{\beta}_R^\top)^\top$ denotes a $K \times 1$ vector containing all regression parameters. Similarly, we let $\boldsymbol{\lambda} = (\rho_1, \ldots, \rho_{R(R-1)/2}, p_1, \ldots, p_R, \boldsymbol{\tau}_1^\top, \ldots, \boldsymbol{\tau}_R^\top)^\top$ be a $Q \times 1$ vector of all dispersion parameters.

To simplify the discussion, let $\mathcal{Y} = (\boldsymbol{Y}_1^\top, \ldots, \boldsymbol{Y}_R^\top)^\top$ be the $NR \times 1$ stacked vector  of the response variable matrix $\mathbf{Y}_{N \times R}$ by columns. Similarly, let $\mathcal{M} = (\boldsymbol{\mu}_1^\top, \ldots, \boldsymbol{\mu}_R^\top)^\top$ be the $NR \times 1$ stacked vector of the expected values matrix $\mathbf{M}_{N \times R}$ by columns. 

We adopt the following quasi-score function for the regression parameters:
\begin{equation*}
\psi_{\boldsymbol{\beta}}(\boldsymbol{\beta}, \boldsymbol{\lambda}) = \boldsymbol{D}^\top \boldsymbol{C}^{-1}(\mathcal{Y} - \mathcal{M}),
\end{equation*}
where $\boldsymbol{D} = \nabla_{\boldsymbol{\beta}} \mathcal{M}$ is an $NR \times K$ matrix, and $ \nabla_{\boldsymbol{\beta}}$ denotes the gradient operator. The $K \times K$ matrix
\begin{equation}
\label{Sbeta}
\mathrm{S}_{\boldsymbol{\beta}} = \mathrm{E}(\nabla_{\boldsymbol{\beta}} \psi_{\boldsymbol{\beta}}) = -\boldsymbol{D}^\top \boldsymbol{C}^{-1} \boldsymbol{D}
\end{equation}
is the {\em sensitivity matrix} of $\psi_{\boldsymbol{\beta}}$ and the $K \times K$ matrix
\begin{equation}
\label{Vb}
\mathrm{V}_{\boldsymbol{\beta}} = \mathrm{Var}(\psi_{\boldsymbol{\beta}}) = \boldsymbol{D}^\top \boldsymbol{C}^{-1} \boldsymbol{D}
\end{equation}
is the {\em variability matrix} of $\psi_{\boldsymbol{\beta}}$. 

Similarly, we adopt the Pearson estimating function, defined by the components
\begin{equation}
\label{Pearson}
\psi_{\boldsymbol{\lambda}_i}( \boldsymbol{\beta}, \boldsymbol{\lambda}) = \mathrm{tr}(W_{\boldsymbol{\lambda}_i}(\boldsymbol{r}^\top\boldsymbol{r} - \boldsymbol{C})) \quad \text{for} \quad i = 1,\ldots,Q,
\end{equation}
where $W_{\boldsymbol{\lambda}_i} = -\partial \boldsymbol{C}^{-1} / \partial \boldsymbol{\lambda}_i$ and $\boldsymbol{r} = \mathcal{Y} - \mathcal{M}$. Details on how to compute these weight matrices are given in Section \ref{DERIVADAS}.

The entry $(i,j)$ of the $Q \times Q$ sensitivity matrix of $\psi_{\boldsymbol{\lambda}}$ is given by,
\begin{equation}
\label{Slambda}
\mathrm{S}_{\boldsymbol{\lambda}_{ij}} = \mathrm{E} \left (  \frac{\partial}{\partial \boldsymbol{\lambda}_i} \psi_{\boldsymbol{\lambda}_j} \right ) = -\mathrm{tr} \left (W_{\boldsymbol{\lambda}_i} \boldsymbol{C} W_{\boldsymbol{\lambda}_j} \boldsymbol{C} \right).
\end{equation} 
We may show using results about characteristic functions of linear and quadratic forms of non-normal variables \citep{Knight:1985}, that the entry $(i,j)$ of the $Q \times Q$ variability matrix of $\psi_{\boldsymbol{\lambda}}$ is given by
\begin{equation}
\label{Vl}
\mathrm{V}_{\boldsymbol{\lambda}_{ij}} = \mathrm{Cov}(\psi_{\boldsymbol{\lambda}_i},\psi_{\boldsymbol{\lambda}_j}) = 2\mathrm{tr}(W_{\boldsymbol{\lambda}_i} \boldsymbol{C} W_{\boldsymbol{\lambda}_j} \boldsymbol{C}) + \sum_{l=1}^{NR} k^{(4)}_l (W_{\boldsymbol{\lambda}_i})_{ll} (W_{\boldsymbol{\lambda}_j})_{ll}, 
\end{equation} 
where $k^{(4)}_l$ denotes the fourth cumulant of $\mathcal{Y}_l$ to be discussed below, see Eq. (\ref{cumulant}). To take into account the covariance between the vectors $\boldsymbol{\beta}$ and $\boldsymbol{\lambda}$, we compute the {\em cross-sensitivity} and {\em -variability} matrices. The entry $(i,j)$ of the $Q \times K$ cross-sensitivity matrix between $\boldsymbol{\beta}$ and $\boldsymbol{\lambda}$ is given by
\begin{equation}
\label{SbetaLambda}
\mathrm{S}_{\boldsymbol{\beta_i \lambda_j}} = \mathrm{E}\left ( \frac{\partial}{\partial \boldsymbol{\lambda_j}} \psi_{\boldsymbol{\beta}_i} \right ) = \boldsymbol{0}.
\end{equation}

In a similar way the entry $(i,j)$ of the $K \times Q$ cross-sensitivity matrix between $\boldsymbol{\lambda}$ and $\boldsymbol{\beta}$ is given by
\begin{equation}
\label{SlambdaBeta}
\mathrm{S}_{\boldsymbol{\lambda_i \beta_j}} = \mathrm{E}\left ( \frac{\partial}{\partial \boldsymbol{\beta}_j} \psi_{\boldsymbol{\lambda}_i} \right ) = - \mathrm{tr}\left (W_{\boldsymbol{\lambda}i} \boldsymbol{C} W_{\boldsymbol{\beta}j} \boldsymbol{C} \right ).
\end{equation}

We can show that the entry $(i,j)$ of the $K\times Q$ cross-variability matrix between $\boldsymbol{\beta}$ and $\boldsymbol{\lambda}$ is given by
\begin{equation}
\label{Vbl}
\mathrm{V}_{\boldsymbol{\lambda_i \beta_j}} = \mathrm{E} \left[\sum_{k=1}^{NR} \sum_{l=1}^{NR} \sum_{m=1}^{NR} W_{\boldsymbol{\lambda_i}}^{(lm)} \boldsymbol{A}^{(j)}_k r_k r_l r_m  \right ],
\end{equation}
where $\boldsymbol{A} = \boldsymbol{D}^T \boldsymbol{C}^{-1}$ and $\boldsymbol{A}^{(j)}$ denotes the $j^{th}$ column of $\boldsymbol{A}$. In a similar way $W_{\boldsymbol{\lambda_i}}^{(lm)}$ denotes the $l m ^{th}$ entry of the matrix $W_{\boldsymbol{\lambda_i}}$. Furthermore, the joint sensitivity matrix of $\psi_{\boldsymbol{\beta}}$ and $\psi_{\boldsymbol{\lambda}}$ is given by
\begin{equation*}
\mathrm{S}_{\boldsymbol{\theta}} = \begin{pmatrix}
\mathrm{S}_{\boldsymbol{\beta}} & \mathrm{S}_{ \boldsymbol{\beta} \boldsymbol{\lambda}} \\ 
\mathrm{S}_{\boldsymbol{\lambda} \boldsymbol{\beta}}  & \mathrm{S}_{\boldsymbol{\lambda}}
\end{pmatrix},
\end{equation*}
whose entries are defined by (\ref{Sbeta}), (\ref{Slambda}), (\ref{SbetaLambda})  and  (\ref{SlambdaBeta}). Finally, the joint variability matrix of $\psi_{\boldsymbol{\beta}}$ and $\psi_{\boldsymbol{\lambda}}$ is given by
\begin{equation*}
\mathrm{V}_{\boldsymbol{\theta}} = \begin{pmatrix}
\mathrm{V}_{\boldsymbol{\beta}} & \mathrm{V}_{ \boldsymbol{\lambda} \boldsymbol{\beta}}^\top \\ 
\mathrm{V}_{\boldsymbol{\lambda} \boldsymbol{\beta}}  & \mathrm{V}_{\boldsymbol{\lambda}}
\end{pmatrix},
\end{equation*}
whose entries are defined by (\ref{Vb}), (\ref{Vl}) and (\ref{Vbl}).

Let $\hat{\boldsymbol{\theta}} = (\hat{\boldsymbol{\beta}}^\top, \hat{\boldsymbol{\lambda}}^\top)^\top$ be the estimating function estimator of $\boldsymbol{\theta}$. Then the asymptotic distribution of $\hat{\boldsymbol{\theta}}$ is
\begin{equation*}
\hat{\boldsymbol{\theta}} \sim \mathrm{N}(\boldsymbol{\theta}, \mathrm{J}_{\boldsymbol{\theta}}^{-1})
\end{equation*} 
where $\mathrm{J}_{\boldsymbol{\theta}}^{-1}$ is the inverse of Godambe information matrix,
\begin{equation*}
\mathrm{J}_{\boldsymbol{\theta}}^{-1} = \mathrm{S}_{\boldsymbol{\theta}}^{-1} \mathrm{V}_{\boldsymbol{\theta}} \mathrm{S}_{\boldsymbol{\theta}}^{-T},
\end{equation*}
where $\mathrm{S}_{\boldsymbol{\theta}}^{-T} = (\mathrm{S}_{\boldsymbol{\theta}}^{-1})^{T}$. 

\citet{Jorgensen:2004} proposed the {\em modified chaser algorithm} to solve the system of equations $\psi_{\boldsymbol{\beta}} = \boldsymbol{0}$ and $\psi_{\boldsymbol{\lambda}} = \boldsymbol{0}$, defined by
\begin{eqnarray}
\label{chaser}
\boldsymbol{\beta}^{(i+1)} &=& \boldsymbol{\beta}^{(i)} - \mathrm{S}_{\boldsymbol{\beta}}^{-1} \psi_{\boldsymbol{\beta}}(\boldsymbol{\beta}^{(i)}, \boldsymbol{\lambda}^{(i)}) \nonumber \\
\boldsymbol{\lambda}^{(i+1)} &=& \boldsymbol{\lambda}^{(i)} - \mathrm{S}_{\boldsymbol{\lambda}}^{-1} \psi_{\boldsymbol{\lambda}}(\boldsymbol{\beta}^{(i+1)}, \boldsymbol{\lambda}^{(i)}).
\end{eqnarray}
The modified chaser algorithm uses the insensitivity property (\ref{SbetaLambda}), which allows us to use two separate equations to update $\boldsymbol{\beta}$ and $\boldsymbol{\lambda}$. This procedure was implemented in \texttt{R} \citep{R:2015} and some generic functions are made available in the supplement material. The modified chaser algorithm is often quite efficient, but it does not have any way to control the 
step length. Thus, based on ideas from \cite{Soren:1991} we propose the { \em reciprocal likelihood algorithm} involving an additional {\em tuning constant} $\alpha$ to control the step length. The reciprocal likelihood algorithm replaces the second equation of (\ref{chaser}) by
\begin{eqnarray}
\label{Modifiedchaser}
\boldsymbol{\lambda}^{(i+1)} &=& \boldsymbol{\lambda}^{(i)} - [ \alpha \psi_{\boldsymbol{\lambda}}(\boldsymbol{\beta}^{(i+1)}, \boldsymbol{\lambda}^{(i)})^T \psi_{\boldsymbol{\lambda}}(\boldsymbol{\beta}^{(i+1)}, \boldsymbol{\lambda}^{(i)}) \mathrm{V}_{\boldsymbol{\lambda}}^{-1}\mathrm{S}_{\boldsymbol{\lambda}} + \mathrm{S}_{\boldsymbol{\lambda}}]^{-1} \psi_{\boldsymbol{\lambda}}(\boldsymbol{\beta}^{(i+1)}, \boldsymbol{\lambda}^{(i)}).
\end{eqnarray}
The strategy for choosing $\alpha$ used in this paper consists of starting the algorithm with $\alpha = 0$, and continuing with $\alpha = 0$ as long as the proposed value of $\boldsymbol{\lambda}^{i+1}$ corresponds to a positive-definite covariance matrix. In the opposite case, we increase the value of $\alpha$ by a small quantity (e.g. $\epsilon = 0.01$) and try again until the covariance matrix becomes positive-definite, after which we return to $\alpha = 0$, corresponding to the modified chaser algorithm. Compared with conventional step length methods, our method is adaptive in the sense that directions where the estimating function is far from zero are being less penalized.   

To compute the variance of the dispersion parameter estimators we used the empirical fourth cumulants, i.e.
\begin{equation}
\label{cumulant}
k^{(4)}_l = (y_l - \hat{\mu}_l)^4 - 3 \hat{\boldsymbol{C}}_{ll}^2.
\end{equation}
The empirical third central moment was computed based on equation (\ref{Vbl}), ignoring the expectation. 
The main advantage of using empirical third and fourth moments is that the resulting method depends on second-moment assumptions only. 
The additional variability induced by the use of empirical moments implies, however, increased variability of the asymptotic covariance of the dispersion parameter estimators, in particular for small sample sizes.

The Pearson estimating function (\ref{Pearson}) is unbiased only if the vector of regression parameters $\boldsymbol{\beta}$ is known. \citet{Jorgensen:2004} proposed a bias-correction for the Pearson estimating function. The $i$th bias-correction term is given by
\begin{equation}
\label{correction}
b_{\boldsymbol{\lambda}i} = - \mathrm{tr}(J_{\boldsymbol{\beta}}^{(\boldsymbol{\lambda}_i)} J_{\boldsymbol{\beta}}^{-1}),
\end{equation}
where $J_{\boldsymbol{\beta}}$ denotes the Godambe information matrix for $\boldsymbol{\beta}$ and $J_{\boldsymbol{\beta}}^{(\boldsymbol{\lambda}_i)} = \partial J_{\boldsymbol{\beta}}/{\partial \boldsymbol{\lambda}_i}$. The corrected Pearson estimating function may be solved using the same algorithm as for the Pearson estimating function. The variability matrix does not depend on the bias-correction term. This is not the case for the sensitivity matrix, but the contribution of the correction term to the sensitivity is so small that it can be ignored.

  \subsection{Derivatives of the covariance matrix}\label{DERIVADAS}
	The key calculation in relation to the fitting algorithm is to compute the derivative of the covariance matrix $\boldsymbol{C}$. In this Section we will provide details of this calculation for the model presented in Eq. (\ref{McGLM}). 
Let $\rho_i$ for $i = 1, \ldots, R(R-1)/2$ denote the correlation parameters. We use the convention to stack the lower triangle of the correlation matrix $\boldsymbol{\Sigma}_b$ by columns. Let $\boldsymbol{p}$ be an $R \times 1$ vector of power parameters. Finally, let $\boldsymbol{\tau}_R$ be a $D \times 1$ vector of dispersion parameters. To denote a specific element we use the notation $\tau_{rd}$ for $d = 0, \ldots, D$ and $r = 1, \ldots, R$.

The weight matrix is defined by 
\begin{equation*}
\label{weights}
W_{\boldsymbol{\lambda}_i} = -\frac{\partial \boldsymbol{C}^{-1}}{\partial \boldsymbol{\lambda}_i} = \boldsymbol{C}^{-1} \frac{\partial \boldsymbol{C}}{\partial \boldsymbol{\lambda}_i} \boldsymbol{C}^{-1}.
\end{equation*}
The partial derivative of $\boldsymbol{C}$ with respect to the element $\rho_i$ is given by
\begin{equation*}
\frac{\partial \boldsymbol{C}}{\partial \rho_i} = \mathrm{Bdiag}(\tilde{\boldsymbol{\Sigma}}_1, \ldots, \tilde{\boldsymbol{\Sigma}}_R)\left( \frac{\partial \boldsymbol{\Sigma}_b}{\partial \rho_i} \otimes \boldsymbol{I}\right ) \mathrm{Bdiag}(\tilde{\boldsymbol{\Sigma}}_1, \ldots, \tilde{\boldsymbol{\Sigma}}_R).
\end{equation*}
Using elementary matrix calculus the partial derivative of $\boldsymbol{C}$ with respect to the element $p_r$ is given by
\begin{eqnarray}
\label{deriv2}
\frac{\partial \boldsymbol{C}}{\partial p_r} =  \mathrm{Bdiag}(\boldsymbol{0}, \ldots, \frac{\partial\tilde{\boldsymbol{\Sigma}}_r}{\partial p_r}, \ldots, \boldsymbol{0})(\boldsymbol{\Sigma}_b \otimes \boldsymbol{I})\mathrm{Bdiag}(\tilde{\boldsymbol{\Sigma}}_1^T, \ldots, \tilde{\boldsymbol{\Sigma}}_R^T) \nonumber \\
+  \mathrm{Bdiag}(\tilde{\boldsymbol{\Sigma}}_1, \ldots, \tilde{\boldsymbol{\Sigma}}_R)(\boldsymbol{\Sigma}_b \otimes \boldsymbol{I}) \mathrm{Bdiag}(\boldsymbol{0}, \ldots, \frac{\partial\tilde{\boldsymbol{\Sigma}}_r^T}{\partial p_r}, \ldots, \boldsymbol{0}).
\end{eqnarray}
Similar equation may be obtained with respect to the elements in the vector $\boldsymbol{\tau}_R$. Given the block diagonal structure of Eq. ($\ref{deriv2}$), it is enough to compute the derivatives $\partial \tilde{\boldsymbol{\Sigma}}_r/\partial p_r$ and insert in Eq. (\ref{deriv2}). Based on the results from \citet{Simo:2013} the partial derivative of $\tilde{\boldsymbol{\Sigma}}_r$ with respect to $p_r$ and $\tau_{rd}$ are given by
\begin{equation*}
\frac{\partial \tilde{\boldsymbol{\Sigma}}_r}{\partial p_r} = \tilde{\boldsymbol{\Sigma}}_r \Phi\left (\tilde{\boldsymbol{\Sigma}}_r^{-1} \frac{\partial \boldsymbol{\Sigma}_r}{\partial p_r}   \tilde{\boldsymbol{\Sigma}}_r^{-1}\right),
\end{equation*}
and
\begin{equation*}
\frac{\partial \tilde{\boldsymbol{\Sigma}}_r}{\partial \tau_{rd}} = \tilde{\boldsymbol{\Sigma}}_r \Phi\left (\tilde{\boldsymbol{\Sigma}}_r^{-1} \frac{\partial \boldsymbol{\Sigma}_r}{\partial \tau_{rd}}   \tilde{\boldsymbol{\Sigma}}_r^{-1}\right),
\end{equation*}
respectively, where the function $\Phi$ returns the lower triangular part of the argument and half of its diagonal. Now, recalling that $\boldsymbol{\Sigma}_r = \mathrm{V}_r(\boldsymbol{\mu};p)^{\frac{1}{2}} \boldsymbol{\Omega}_r(\boldsymbol{\tau}_r) \mathrm{V}_r(\boldsymbol{\mu};p)^{\frac{1}{2}}$, we may hence see that the partial derivative with respect to $p_r$ and $\tau_{rd}$ are given by

\begin{equation}
\label{DevVar}
\frac{\partial \boldsymbol{\Sigma}_r}{\partial p_r} = \frac{\partial \mathrm{V}_r(\boldsymbol{\mu};p)^{\frac{1}{2}}}{\partial p_r}\boldsymbol{\Omega}_r(\boldsymbol{\tau}_r) \mathrm{V}_r(\boldsymbol{\mu};p)^{\frac{1}{2}} + \mathrm{V}_r(\boldsymbol{\mu};p)^{\frac{1}{2}} \boldsymbol{\Omega}_r(\boldsymbol{\tau}_r)\frac{\partial \mathrm{V}_r(\boldsymbol{\mu};p)^{\frac{1}{2}}}{\partial p_r},
\end{equation}
and
\begin{equation*}
\frac{\partial \boldsymbol{\Sigma}_r}{\partial \tau_{rd}} = \mathrm{V}_r(\boldsymbol{\mu};p)^{\frac{1}{2}} \frac{\partial \boldsymbol{\Omega}_r(\boldsymbol{\tau}_r)}{\partial \tau_{rd}} \mathrm{V}_r(\boldsymbol{\mu};p)^{\frac{1}{2}},
\end{equation*}
respectively, where
\begin{equation}
\label{devOmega}
\frac{\partial \boldsymbol{\Omega}_r(\boldsymbol{\tau}_r)}{\partial \tau_{rd}} = \frac{\partial h^{-1}(\boldsymbol{U})}{\partial \boldsymbol{U}} Z_{rd}
\end{equation}
and where $\boldsymbol{U} = \tau_{r0}Z_{r0}+\cdots+\tau_{rD}Z_{rD}$. The derivative in Eqs. (\ref{DevVar}) and (\ref{devOmega}) depends on the derivative of the variance function and covariance link function, respectively, and it should be evaluated accordingly.

\section{Data analyses}\label{results}
\subsection{Results from data set 1}
	In this section we apply the McGLM approach to analyze the multivariate count data set presented in Section \ref{data1}. We adopted the log link function, the Poisson-Tweedie variance function and the identity covariance link function for the five count response variables. The matrix linear predictor is composed of an identity matrix, since we have independent respondents. The linear predictor is composed of nine covariates plus the intercept for each response variable. The covariance structure is described by five power parameters, five dispersions and ten correlation parameters. We fitted this model using the \textit{modified chaser} algorithm (\ref{chaser}) and the correction term in (\ref{correction}). Table \ref{tab:exemplo1} shows the estimates and standard errors for the power and dispersion parameters.

\begin{table}
\caption{\label{tab:exemplo1}Power and dispersion parameter estimates and standard errors (SE) for the Australian healt survey data.}
\centering
\fbox{%
\begin{tabular}{l|cc|cc|cc|cc|cc}
 & \multicolumn{2}{c}{\texttt{Ndoc}} & \multicolumn{2}{c}{\texttt{Nndoc}} & \multicolumn{2}{c}{\texttt{Nmed}} & \multicolumn{2}{c}{\texttt{Nhosp}} & \multicolumn{2}{c}{\texttt{Nadm}} \\ 
\hline
               & Estimate   & SE   & Estimate & SE      & Estimate & SE      & Estimate  & SE     & Estimate  & SE    \\
     $\hat{p}$ & 1.1414  & 0.3513  & 1.1345   & 0.2638  & 1.2653   & 0.2830  & 1.6394    & 0.1510 & 1.7419    & 0.5783        \\
$\hat{\tau}_0$ & 1.0820  & 0.4352  & 3.3761   & 1.2598  & 0.3570   & 0.0646  & 18.2059   & 3.2348 & 1.2065    & 0.9371        \\ \hline
\end{tabular}}
\end{table}

\begin{figure}[htbp]
\centering
\includegraphics{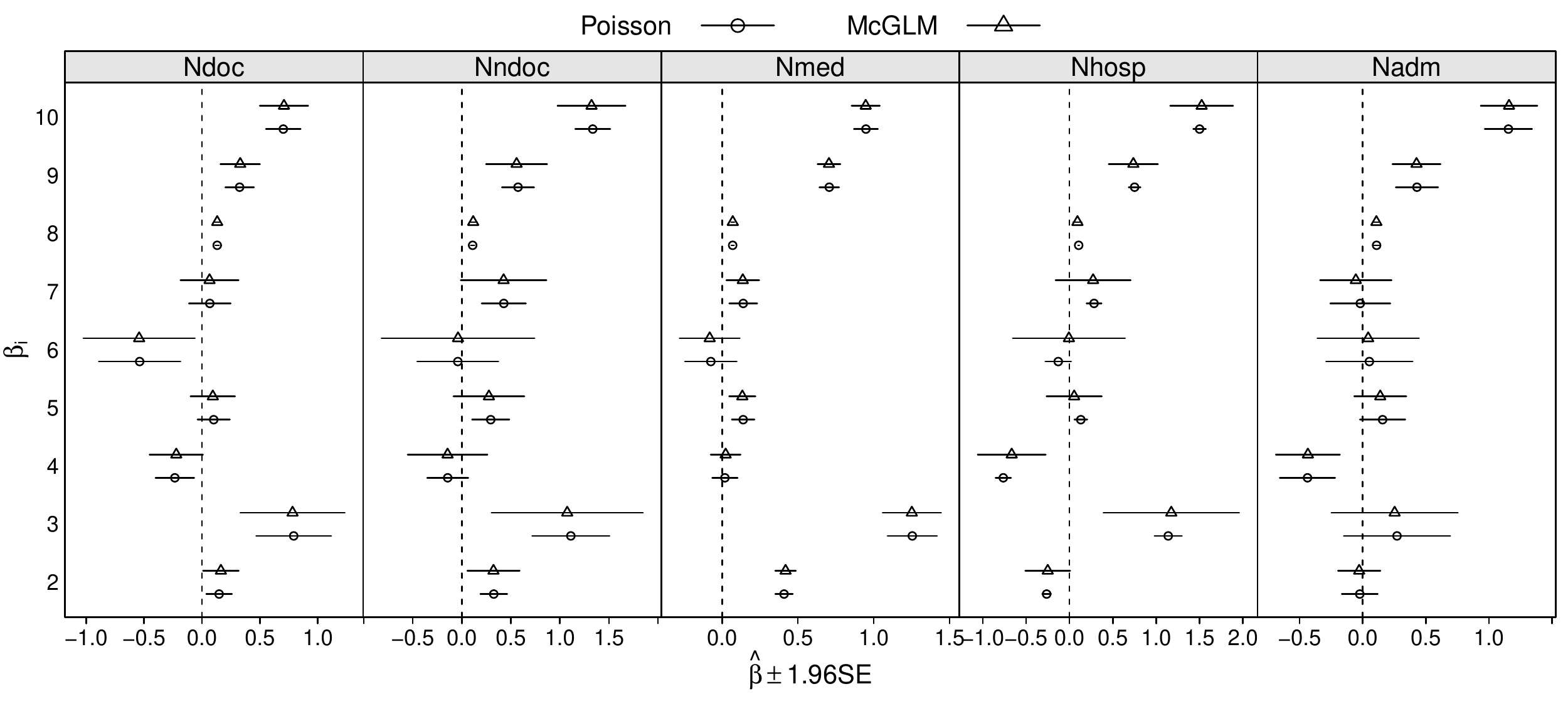}
\caption{Regression parameter estimates and $95\%$ confidence intervals by response variable and model for the Australian healt survey data.}
\label{fig:exemplo1}
\end{figure}

The results in Table \ref{tab:exemplo1} show that for the response variables \texttt{Ndoc}, \texttt{Nndoc} and \texttt{Nmed} the suggested distribution is the Neyman Type A ($p=1$), which indicates zero inflation relative to the Poisson distribution. Regarding the response variable \texttt{Nhosp} the model indicates that the P\'{o}lya-Aeppli distribution ($p=1.5$) is suitable. Finally, the model indicates that for \texttt{Nadm} both Neyman Type A, P\'{o}lya-Aeppli and negative binomial ($p=2$) distributions are suitable. This result is obtained because the dispersion parameter $\tau_0$ is not different from zero; hence the response variable \texttt{Nadm} is equidispersed and all these distributions work well, including the Poisson. In this case, we do not have enough information in the data to distinguish between these distributions. Therefore, we suggest to opt for the simplest possibility, i.e. the Poisson model. The dispersion estimates show weak overdispersion for the response variables \texttt{Ndoc}, \texttt{Nndoc} and \texttt{Nmed} and high overdispersion for \texttt{Nhosp}. In order to compare the regression coefficients with a conventional model, Figure \ref{fig:exemplo1} shows the estimates and confidence intervals for McGLM and a conventional Poisson log-linear model for each response variable. The intercept is not shown in order to avoid scale issues.

\setkeys{Gin}{width=0.99\textwidth}

The results in Figure \ref{fig:exemplo1} show that the two approaches agree in terms of estimates, but differ in terms of standard errors. The differences may be explained by the covariance structure. The Poisson model assumes equidispersion, whereas the McGLM models allow for a flexible modelling of the covariance structure, allowing in particular various degrees of overdispersion and zero-inflation. For the response variable \texttt{Nadm}, the model shows that equidispersion is suitable, making the McGLM and Poisson confidence intervals similar. On the other hand, for the response variable \texttt{Nhosp}, where the overdispersion is strong, the McGLM confidence intervals  are about five times wider than the Poisson ones. In a similar way, the McGLM confidence intervals for \texttt{Nndoc}, \texttt{Ndoc} and \texttt{Nmed} are on average $93\%$, $38\%$ and $17\%$ wider than the corresponding Poisson intervals. These results highlight the importance of modelling the covariance structure even when the main interest is in the regression parameters, because the covariance structure controls the standard errors for the regression parameters.   

An additional feature of McGLM is that we can estimate the correlation between response variables. It is important to emphasize that the estimation of the correlation matrix does not inflate the standard errors for the regression coefficients, due to the insensitivity of the quasi-score function with respect to the covariance parameters. The estimates and standard errors for the entries of the $\boldsymbol{\Sigma}_b$ matrix were as follows:
$$
\hat{\boldsymbol{\Sigma}}_b = \begin{bmatrix}
 1 &  &  &  & \\ 
0.1066(0.0161) & 1 &  &  & \\ 
0.1708(0.0156) & 0.0601(0.0144) & 1 &  & \\ 
0.0905(0.0164) & 0.0679(0.0156) & 0.0478(0.0144) & 1 & \\ 
0.1503(0.0160) & 0.0688(0.0147) & 0.0699(0.0140) & 0.5464(0.0510) & 1 
\end{bmatrix}.
$$
All correlations are significantly different from zero, but only the correlation between \texttt{Nhosp} and \texttt{Nadm} is substantial in size. The standard errors are all of a similar magnitude, which is natural since all are computed using the same sample size. Furthermore, these correlations take into account the effect of all covariates, zero inflation and overdispersion. We know of no other statistical method that allows estimation of correlations taking into account all these important features.

\subsection{Results from data set 2}
	In this section, we apply the McGLM approach to analyze data set 2 from Section \ref{data2}, which has response variables of mixed types. There are three response variables, namely \texttt{HR}, \texttt{RR} and \texttt{O}$_2$\texttt{Sat}, the first two being continuous and the last being confined to the unit interval, having exact zeroes. 
We adopted the constant variance function, identity link function, and identity covariance link function for \texttt{HR} and \texttt{RR}, reflecting a belief that \texttt{HR} and \texttt{RR} are normally distributed. For \texttt{O}$_2$\texttt{Sat} we adopted the logit link function combined with the binomial variance function and identity covariance link function.  We fitted the model using the modified chaser algorithm (\ref{chaser}) and the correction term (\ref{correction}). 

The matrix linear predictor is composed of a diagonal matrix (intercept) combined with two sets of matrices to model the longitudinal and repeated measures structures. The longitudinal structure is modeled by a compound symmetry matrix (of ones), the reciprocal of Euclidean distances and reciprocal of Euclidean distances squared. The repeated measures structure is described by an unstructured covariance matrix. Since we have three \texttt{Evaluations} to represent this structure we need three matrices. Therefore, the matrix linear predictor is a linear combination of seven known matrices and it is described by $21$ dispersion parameters (seven for each outcome). Details of the matrix linear predictor is available in the supplementary material. In this example we have no power parameters and the $\boldsymbol{\Sigma}_b$ matrix contains three parameters. Table \ref{tab:exemplo2} shows the estimates and standard errors for the dispersion parameters. The parameters ($\tau_1,\tau_2$ and $\tau_3$) and ($\tau_4,\tau_5$ and $\tau_6$) are associated to the repeated measures and longitudinal structures, respectively.

\begin{table}
\caption{\label{tab:exemplo2}Dispersion parameter estimates and standard errors (SE) for the Respiratory physiotherapy on premature newborns data.}
\centering
\fbox{%
\begin{tabular}{l|cc|cc|cc}
 & \multicolumn{2}{c}{\texttt{RR}} & \multicolumn{2}{c}{\texttt{HR}} & \multicolumn{2}{c}{\texttt{O}$_2$\texttt{Sat}} \\ 
\hline
                    & Estimate   & SE        & Estimate  & SE         & Estimate& SE          \\
     $\hat{\tau}_0$ & 134.2669   & 24.2386   & 95.2933   & 32.3082    & 0.0129   & 0.0840         \\
     $\hat{\tau}_1$ & 40.6929    & 11.6902   & 82.8065   & 15.3124    & 0.0010   & 0.0051          \\
     $\hat{\tau}_2$ & 15.5255    & 12.0365   & 56.8178   & 15.6080    & 0.0036   & 0.0364          \\
     $\hat{\tau}_3$ & 50.2033    & 12.6795   & 54.2204   & 15.5270    & 0.0012   & 0.0064          \\
     $\hat{\tau}_4$ & $-$18.6080   & 19.8331   & 39.9998   & 24.2276    & $-$0.0003  & 0.0172           \\
     $\hat{\tau}_5$ & 81.1339    & 75.3181   & $-$143.9326  & 91.3125   & $-$0.0001  & 0.0534           \\
     $\hat{\tau}_6$ & $-$63.0717   & 57.3648   & 105.7698  & 68.6823    & $-$0.0017  & 0.0375           \\ \hline
\end{tabular}}
\end{table}

The results in Table \ref{tab:exemplo2} show that the longitudinal structure is not significant for all response variables. The repeated measures structure is significant for \texttt{RR} and \texttt{HR}. For the outcome \texttt{RR} the estimate of $\tau_2$ is not significant, which means that the covariance between \texttt{Evaluation 1} and \texttt{Evaluation 3} is not different from zero. For the outcome \texttt{O}$_2$\texttt{Sat} there are no significant dispersion coefficients, so we may assume independent observations. The final model is composed of the repeated measures structure for the response variables \texttt{RR} and \texttt{HR} and independent structure (only $\tau_0$) for \texttt{O}$_2$\texttt{Sat}. 

We have a set of $16$ covariates entering the linear predictor, and we used a stepwise procedure to select the most significant set of covariates. This procedure selected a different set of covariates for each outcome. After completing this procedure, we included the covariate of particular interest, namely \texttt{treat}, which is a factor with two levels. Our goal is to assess whether or not the treatment has an effect on each response variable. 

\begin{figure}[htbp]
\centering
\includegraphics{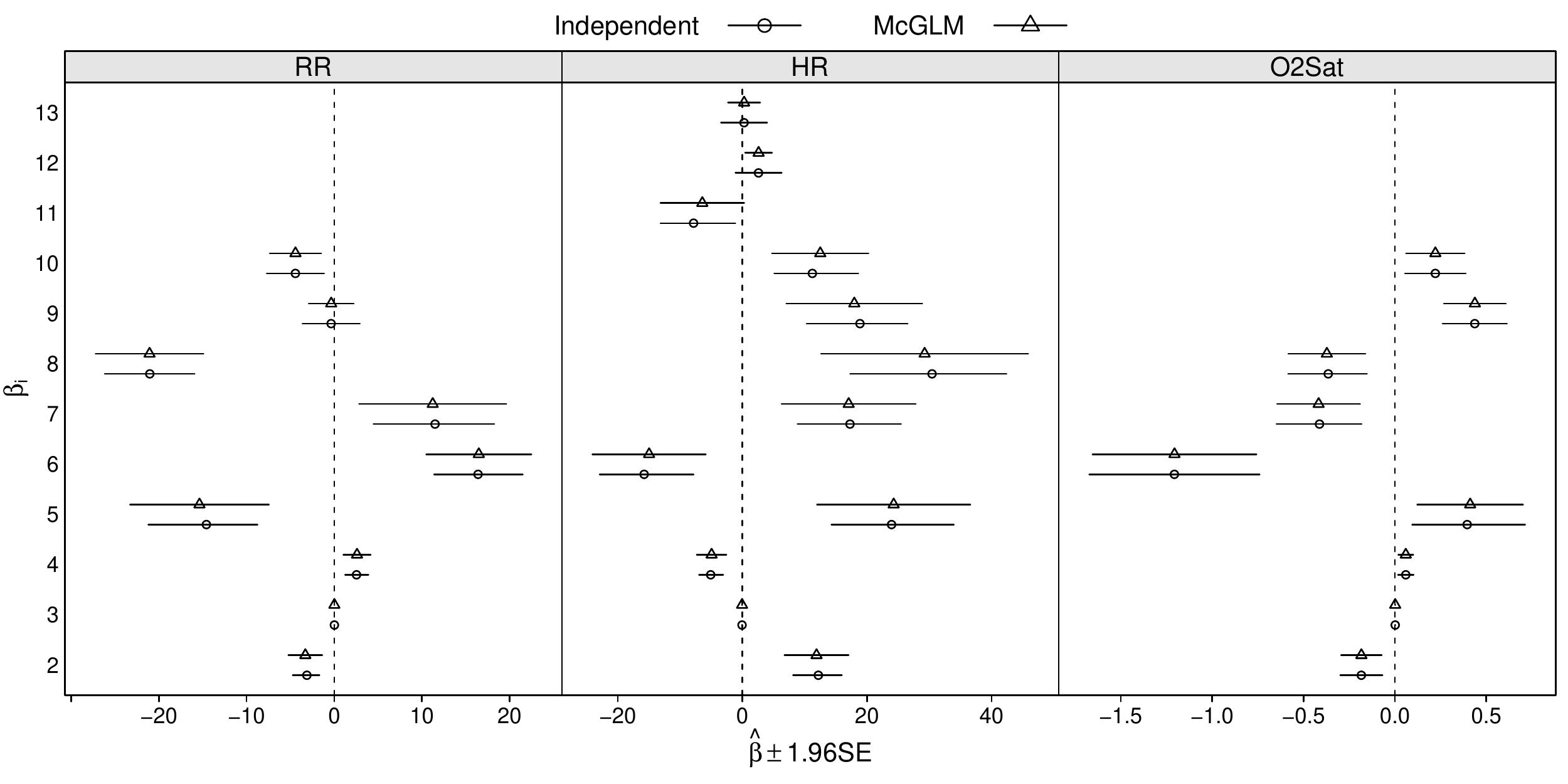}
\caption{Regression paratemeter estimates and $95\%$ confidence intervals by outcome and model for the Respiratory physioterapy data set.}
\label{fig:exemplo2}
\end{figure}

In order to evaluate the effects of the covariance structure on the regression coefficients, Figure \ref{fig:exemplo2} shows estimates and confidence intervals obtained from the final McGLM and a quasi GLM using the same link and variance functions as for the McGLM. The linear predictors for the outcome \texttt{RR}, \texttt{HR} and \texttt{O}$_2$\texttt{Sat} are composed of $10$, $13$ and $10$ regression coefficients, respectively. The intercept is not shown, in order to avoid scaling issues. It is important to emphasize that the last two regression coefficients (numbered 9--10, 12--13 and 9--10, respectively) measure the treatment effects.

\setkeys{Gin}{width=0.99\textwidth}

The results in Figure \ref{fig:exemplo2} show that in general the confidence intervals from McGLM are wider than the corresponding ones based on quasi GLM. For the outcome \texttt{RR} and \texttt{HR} the standard errors from McGLM are on average $19\%$ and $27\%$ greater than the corresponding quasi GLM ones, respectively. These results are as expected, because correlation within response variables generally implies less information in the data on the regression coefficients. It is hence interesting to note that, in contras to the other regression coefficients, the two treatment coefficients for each response variable have smaller standard errors using McGLM than those obtained using quasi GLM. We attribute this effect to the fact that the treatment covariate is also used for modelling the covariance structure, which apparently improves on the estimation of the treatment effect, although we are uncertain if this is a general feature, or if it is specific to this data set. 

Regarding the outcome \texttt{O}$_2$\texttt{Sat} the standard errors from McGLM are on average $4\%$ smaller than those obtained by quasi GLM. This may be explained by the difference between the estimates of $\tau_0$, which are $0.0129$ and $0.0138$ for McGLM and quasi binomial, respectively. This small difference seems to be due to the use of the corrected Pearson estimator in our model. 

Regarding the treatment effects, the final model shows that for \texttt{RR} the \texttt{Evaluation 1} differs from \texttt{Evaluation 3}, but not from \texttt{Evaluation 2}. On the other hand, for \texttt{HR} the model shows that the \texttt{Evaluation 1} differs from \texttt{Evaluation 2}, but does not differ from \texttt{Evaluation 3}. This result contrasts with the quasi GLM analysis, which does not show any significant difference between \texttt{Evaluation 1} and \texttt{Evaluation 2}. Finally, our final model shows that for \texttt{O}$_2$\texttt{Sat} both constrasts are significant. The estimated correlation matrix between response variables was as follows: 
$$
\hat{\boldsymbol{\Sigma}}_b = \begin{bmatrix}
1 &  & \\ 
0.1682 (0.0607) & 1 & \\ 
$-$0.0482 (0.0607) & $-$0.0733 (0.0608) & 1
\end{bmatrix}
$$
We observe that there is a significant but weak correlation between \texttt{RR} and \texttt{HR}, whereas the other two correlation estimates are not significant.

\subsection{Results from data set 3}\label{results3}
	In this section we apply the McGLM approach to the space-time data set presented in Section \ref{data3}. The response variable monthly rainfall is right-skewed with a positive probability at zero. We hence adopted the log link function, the power variance function and the inverse covariance link function. The linear predictor is expressed in terms of Fourier harmonics (seasonal variation) and B-splines (general trend),
\begin{equation*}
g(\mu_{tj}) = \beta_0 + \beta_1 \mathrm{cos}(2\pi t/12) + \beta_2 \mathrm{sin}(2\pi t /12) + \sum_{k=1}^4\beta_{k+2}B_{k}(j)
\end{equation*}
where $t = 1, \ldots,192$ indexes months, and $j = 1, \ldots, 16$ indexes years. The $B_{k}(j)$ form a B-spline basis with four degree of freedom. We used only the first term of the Fourier harmonics, since the second and third terms were not significant.

The main challenge in the analysis was to model the covariance structure suitably, in order to take into account the spatial and temporal autocorrelation and, if necessary, the interaction between space and time. We propose to model the space-time structure using a linear combination of neighborhood matrices. Let us motivate our approach using the Conditional Autoregressive (CAR) model. The CAR model specifies the inverse of the covariance matrix by
\begin{equation*}
\boldsymbol{\Omega}^{-1}(\tau, \rho) = \tau (\boldsymbol{D} - \rho \boldsymbol{W})
\end{equation*}
where $\boldsymbol{W}$ is a neighborhood matrix and $\boldsymbol{D}$ is a diagonal matrix with the number of neighbors in the main diagonal. The model is parametrized by the precision ($\tau$) and autocorrelation ($\rho$) parameters. The matrices $\boldsymbol{D}$ and $\boldsymbol{W}$ can model both space, time and space-time interaction in a straightforward way, by using different neighborhood matrices. 

For the Venezuelan rainfall data, we used the spatial coordinates (\texttt{latitude} and \texttt{longitude}) to build a Voronoi tessellation (see Figure \ref{fig:exploratory3}B). The tessellation structure helps us to specify a neighborhood structure. Let us denote this structure by $\boldsymbol{D}_s$ and $\boldsymbol{W}_s$. Temporal neighbors are naturally specified by the time structure. Let us denote these matrices by $\boldsymbol{D}_t$ and $\boldsymbol{W}_t$. In the space-time case we have replicates of the space and time structures, so assuming independent replicates, the full neighborhood matrix is block-diagonal. Finally, the interaction between space and time is described by the Kronecker product between the space and time neighborhood structures. Let us denote these matrices by $\boldsymbol{D}_{st} = \boldsymbol{D}_{s} \otimes \boldsymbol{D}_{t}$ and $\boldsymbol{W}_{st} = \boldsymbol{W}_{s} \otimes \boldsymbol{W}_{t}$. Thus, the matrix linear predictor is given by
\begin{equation*}
\boldsymbol{\Omega}^{-1}(\boldsymbol{\tau}, \boldsymbol{\rho}) = \tau_t(\boldsymbol{D}_t + \rho_t \boldsymbol{W}_t) + \tau_s(\boldsymbol{D}_s + \rho_s \boldsymbol{W}_s) + \tau_{st}(\boldsymbol{D}_{st} + \rho_{st} \boldsymbol{W}_{st}), 
\end{equation*}
which may be written as a linear combination of known matrices as follows:
\begin{equation}
  \label{eq:COVEX3}
\boldsymbol{\Omega}^{-1}(\boldsymbol{\tau}) = \tau_0 Z_0 + \tau_1 Z_1 + \tau_2 Z_2 + \tau_3 Z_3 +\tau_4 Z_4 +\tau_5 Z_5,
\end{equation}
where $Z_0 = \boldsymbol{D}_t$, $Z_1 = \boldsymbol{W}_t$, $Z_2 = \boldsymbol{D}_s$, $Z_3 = \boldsymbol{W}_s$, $Z_4 = \boldsymbol{D}_{st}$ and $Z_5 = \boldsymbol{W}_{st}$. In a similar way, we find that $\rho_t = \tau_1/\tau_0$, $\rho_s = \tau_3/ \tau_2$ and $\rho_{st} = \tau_5/\tau_4$. Finally, $\tau_t = \tau_0$, $\tau_s = \tau_2$ and $\tau_{st} = \tau_4$. In practical situations, fitting this model may be hard if the autocorrelation parameters are near $1$. In the Bayesian context, it is common to fix the autocorrelation parameter $\rho$ at the value 1, which is the so-called Intrinsic Conditional Autoregressive (ICAR) model.

In order to investigate the space-time structure we fitted three models, cf. Table \ref{tab:exemplo3}. The first (\texttt{Model 1}) considers time effects. The second (\texttt{Model 2}) considers space effects and the third (\texttt{Model 3}) the space-time interaction effects. After this procedure we decided that the autocorrelation parameters associated with space and interaction effects must be fixed at the value $1$. We then fitted the final model (\texttt{Model 4}) with all three components time, space and space-time interaction. All models were fitted using the reciprocal likelihood algorithm (\ref{Modifiedchaser}). Table \ref{tab:exemplo3} presents estimates and standard errors for the power and dispersion parameters for each of the four models.  

\begin{table}
\caption{\label{tab:exemplo3}Covariance, power and dispersion parameter estimates and standard errors (SE) for each model for the Venezuelan rainfall data.}
\centering
\fbox{%
\begin{tabular}{l|cc|cc|cc|cc}
\hline
& \multicolumn{2}{c}{\texttt{Model 1}} & \multicolumn{2}{c}{\texttt{Model 2}} & \multicolumn{2}{c}{\texttt{Model 3}} & \multicolumn{2}{c}{\texttt{Model 4}} \\ \hline
                & Estimate    & SE      & Estimate    & SE       & Estimate    & SE       & Estimate & SE    \\ 
$\hat{p}$       & 1.2731      & 0.0380  & 1.1241      & 0.1926   & 1.2340      & 0.1066   & 1.1443  & 0.2475        \\ 
$\hat{\tau}_0$  & 1.1505      & 0.0349  & -           & -        & -           & -        & 0.0339  & 0.0074        \\ 
$\hat{\tau}_1$  & 0.5285      & 0.0268  & -           & -        & -           & -        & 0.0264  & 0.0051        \\ 
$\hat{\tau}_2$  & -           & -       & 0.9252      & 0.2175   & -           & -        & 0.6084  & 0.2844        \\ 
$\hat{\tau}_3$  & -           & -       & 0.9169      & 0.2185   & -           & -        & -       & -             \\ 
$\hat{\tau}_4$  & -           & -       & -           & -        & 0.4487      & 0.0390   & 0.1791  & 0.0333        \\ 
$\hat{\tau}_5$  & -           & -       & -           & -        & 0.3917      & 0.0399   & -       & -             \\ \hline
$\hat{\rho}$    & 0.4593      & 0.1079  & 0.9910      & 0.0038   & 0.8729      & 0.0159   & 0.7787  & 0.0998       
\end{tabular}}
\end{table}

The results in Table \ref{tab:exemplo3} show a moderate temporal autocorrelation, but high spatial and space-time interaction autocorrelations. All the power parameter estimates are in the interval $1 \leq p \leq 2$, suggesting a compound Poisson distribution, as expected, since the response variable is continuous with exact zeros. The fitted values and $95\%$ confidence interval are shown in Figure \ref{fig:exploratory3}A above.

The model allows us to make prediction using the Best Linear Unbiased Predictor (BLUP). Furthermore, we may obtain predictions for different response variable measures, such as the mean number of precipitations events per month, the avarage amount of precipitation per event, or the probability of no precipitation. Such extensions are straightforward and will be presented elsewhere. 

\section{Discussion}\label{discussion}
	In this paper we have developed a comprehensive statistical modelling framework for correlated data, obtained by using separate pairs of link functions and linear predictors for the mean and covariance structures in the style of Pourahmadi (2011). Motivated by three data examples, we have shown that the McGLM framework can deal with a wide variety of correlation structures where existing modelling approaches have difficulties. Following Nelder and Wedderburn (1972), there are obvious pedagogical advantages to the modular specification of models in McGLM, incentivating the researcher to think constructively about the covariance structure, while drawing on previous experiences from GLM modelling. The generalized Kronecker product facilitates the specification of the covariance structure for multivariate response variables.

The main features of the McGLM framework include the ability to deal with most common types of response variables, such as continuous, count and binary. Characteristics such as symmetry/asymmetry, excess of zeros and overdispersion are easily handled due to the flexibility of the model class. We can model many different types of dependencies, such as repeated measures, longitudinal, time series, spatial and space-time data. All of these features extend to multivariate response variables, including the case of mixed response variable types, while maintaining the \textit{population average} interpretations of regression parameters as well as for covariance parameters. This gives a very flexible modelling of the covariance structure compared with for example current GEE implementations, where the researcher must choose from a small set of pre-defined covariance models.  

The main technical advantage of the McGLM framework is the simplicity of the fitting method, which amounts to finding the root for a set of non-linear equations. Based on second-moment assumptions, we use a quasi-score function for the regression parameters and a Pearson estimating function for the covariance parameters. The modified chaser algorithm of Jørgensen and Knudsen (2004) requires an approximate derivative matrix in the form of the sensitivity matrix, which is usually relatively simple. The new reciprocal likelihood algorithm requires the additional calculation of the variability matrix in order to stabilize the covariance parameter update, resulting in a very efficient algorithm, although the sensitivity matrix may be hard to calculate for big data. In such cases a numerical approximation for the sensitivity matrix may be used. For both algorithms a careful choice of initial values is required.
In any case, we avoid using computationally more intensive methods such as MCMC, numerical optimization or numerical integration, which are common in the context of random effects models. 

An important feature of the McGLM fitting method is that while the mean parameter estimators depend relatively little on the form of the covariance structure, this is not the case for the standard errors of the mean parameter estimators, which depend directly on the choice of covariance structure. A related matter is that the discussion of the efficiency of the mean and covariance parameter estimators is difficult due to the lack of a fully specified probability model. 

The current version of the fitting algorithm (available in the supplementary material) is a preliminary implementation of the McGLM method. We plan to develop a full McGLM \texttt{R} package with a GLM-style interface that takes full advantage of the modular specification of the models. There are many possible extensions to the basic model discussed in the present paper, including for example facilities for censored data in survival analysis and other special types of data, or new estimating functions to handle data not missing at random. The specification of the matrix linear predictor is one of the key points of the McGLM approach. While we have used some simple and easily interpretable matrices here, there is wide scope for further research on the proper choice of the matrix linear predictor. Similarly, other covariance link functions, such as the matrix-logarithm \citet{Chiu:1996} may also be explored. It is also possible to incorporate penalized splines into the mean and covariance structures, and to use regularization for high-dimensional data, with important applications in genetics. Furthermore, McGLMs may be scaled to test for a common exposure effect in the style of \citet{Roy2003}.

\section*{Appendix - Data sets description}
	\textbf{Data set 1: Australian health survey} \\
\textbf{Response variables}: 
\texttt{Ndoc} - Number of consultations with a doctor or specialist.
\texttt{Nndoc} - Number of consultations with health professionals.
\texttt{Nmed} - Total number of prescribed and non prescribed medications used in the past two days.
\texttt{Nhosp} - Number of nights in a hospital during the most recent admission.
\texttt{Nadm} - Number of admissions to a hospital, psychiatric hospital, nursing or convalescence home in the past 12 months.\\
\textbf{Covariates}:
\texttt{sex} - factor, two levels (0-Male; 1-Female).
\texttt{age} - respondent's age in years divided by 100.
\texttt{income} - respondent's annual income in Australian dollars divided by 1000.
\texttt{levyplus} - factor, two levels (1- if respondent is covered by private health insurance fund for private patients in public \texttt{hospital} (with doctor of choice); 0 - otherwise).
\texttt{freepoor} - factor, two levels (1 - if respondent is covered by government because low income, recent immigrant, unemployed; 0 - otherwise).
\texttt{freerepa} - factor, two levels (1 - if respondent is covered free by government because of old-age or disability pension, 
or because invalid veteran or family of deceased veteran; 0 - otherwise).
\texttt{illness} - number of illnesses in past 2 weeks, with 5 or more weeks coded as 5.
\texttt{actdays} - number of days of reduced activity in the past two weeks due to illness or injury.
\texttt{hscore} - respondent's general health questionnaire score using Goldberg's method; high score indicates poor health.
\texttt{chcond1} - factor, two levels (1 - if respondent has chronic condition(s) but is not limited in activity; 0 - otherwise).
\texttt{chcond2} - factor, two levels (1 if respondent has chronic condition(s) and is limited in activity; 0 - otherwise).
id - respondent's index.\\
\textbf{Data set 2: Respiratory physiotherapy on premature newborns}\\
\textbf{Response variables}:
\texttt{RR} - Respiratory rate (continuous).
\texttt{HR} - Heart rate (continuous).
\texttt{O2Sat} - Oxygen saturation (bounded).\\
\textbf{Covariates}:
\texttt{Sex} - factor, two levels (Female; Male).
\texttt{GA} - Gestational age (weeks).
\texttt{BW} - Birth weight (mm).
\texttt{APGAR1M} - APGAR index in the first minute of life.
\texttt{APGAR5M} - APGAR index in the fifth minute of life.
\texttt{PRE} - factor, two levels (Premature: yes; no).
\texttt{HD} - factor, two levels (Hansen's disease, yes; no).
\texttt{SUR} - factor, two levels (Surfactant, yes; no).
\texttt{JAU} - factor, two levels (Jaundice, yes; no).
\texttt{PNE} - factor, two levels (Pneumonia, yes; no).
\texttt{PDA} - factor, two levels (Persistence of ductus arteriosus, yes; no).
\texttt{PPI} - factor, two levels (Primary pulmonary infection, yes; no).
\texttt{OTHERS} - factor, two levels (Other diseases, yes; no).
\texttt{DAYS} - Age (days).
\texttt{AUX} - factor, two levels (Type of respiratory auxiliary, HOOD; OTHERS).
\texttt{TREAT} - factor, three levels (Respiratory physiotherapy, Evaluation 1; Evaluation 2; Evaluation 3).
\texttt{UNIT} - Unit sample code.
\texttt{TIME} - Days of treatment. \\
\textbf{Data set 3: Venezuelan rainfall data}\\
\textbf{Response variable}:
\texttt{rainfall} - monthly rainfall (mm)\\
\textbf{Covariates}:
\texttt{month} - month code.
\texttt{Longitude} - Longitude (UTM).
\texttt{Latitude} - Latitude (UTM).
\texttt{height} - height above sea level (m). \\
\textbf{Supplement material web page}

\texttt{http://www.leg.ufpr.br/doku.php/publications:papercompanions:mcglm}

\section*{Acknowledgements}
The first author is supported by CAPES (Coordena\c{c}\~ao de Aperfei\c{c}oamento de Pessoal de N\'ivel Superior) - Brazil.

\bibliographystyle{dcu}
\bibliography{BonatJorgensen2015}
\end{document}